\documentclass[referee]{aa}

\usepackage{graphicx}
\usepackage{natbib}
\usepackage{caption}
\usepackage{subcaption}
\usepackage{multirow}

\usepackage{lineno}
\usepackage{txfonts}

\usepackage[breaklinks=true]{hyperref} 
\begin{document}

   \title{Coronal Diagnostics of Solar Type-III Radio Bursts Using LOFAR and PSP Observations}
   
\titlerunning{LOFAR Type III Imaging}


   \author{Mohamed Nedal\inst{1}\fnmsep\thanks{Corresponding author}
          Kamen Kozarev\inst{1}
          Peijin Zhang\inst{1,3}
          \and
          Pietro Zucca\inst{2}
          }

   \institute{Institute of Astronomy of the Bulgarian Academy of Sciences, Sofia 1784, Bulgaria\\
             \email{mnedal@nao-rozhen.org}
         \and
             ASTRON Netherlands Institute for Radio Astronomy, Oude Hoogeveensedijk 4, 7991 PD, Dwingeloo 7990 AA, The Netherlands
             \and
             Department of Physics, University of Helsinki, P.O. Box 4, Yliopistonkatu 3, Helsinki 00014, Finland
             }


 
  \abstract
   {Solar type III radio bursts are quite common phenomena. They are the result of accelerated electron beams propagating through the solar corona. These bursts are of particular interest as they provide valuable information about the magnetic field and plasma conditions in the corona, which are difficult to measure directly.
   }
   {This study aims to investigate the ambiguous source and the underlying physical processes of the type III radio bursts that occurred on April 3, 2019, through the utilization of multi-wavelength observations from the Low-Frequency Array (LOFAR) radio telescope and the Parker Solar Probe (PSP) space mission, as well as incorporating results from a Potential Field Source Surface (PFSS) and magnetohydrodynamic (MHD) models. The primary goal is to identify the spatial and temporal characteristics of the radio sources, as well as the plasma conditions along their trajectory.}
   {Data preprocessing techniques are applied to combine high- and low-frequency observations from LOFAR and PSP between 2.6 kHz and 80 MHz. We then extract information on the frequency drift and speed of the accelerated electron beams from the dynamic spectra. Additionally, we use LOFAR interferometric observations to image the sources of the radio emission at multiple frequencies and determine their locations and kinematics in the corona. Lastly, we analyze the plasma parameters and magnetic field along the trajectories of the radio sources using PFSS and MHD model results.}
   {We present several notable findings related to type III radio bursts. Firstly, through our automated implementation, we were able to effectively identify and characterize 9 type III radio bursts in the LOFAR-PSP combined dynamic spectrum and 16 type III bursts in the LOFAR dynamic spectrum. We found that the frequency drift for the detected type III bursts in the combined spectrum is ranging between 0.24 and 4 MHz s$^{-1}$, and the speeds of the electron beams are ranging between 0.013 and 0.12 C. Secondly, our imaging observations show that the electrons responsible for these bursts originate from the same source and within a short time frame of fewer than 30 minutes. Finally, our analysis provides informative insights into the physical conditions along the path of the electron beams. For instance, we found that the plasma density obtained from the Magnetohydrodynamic Algorithm outside a Sphere (MAS) model is significantly lower than the expected theoretical density.}
   {}


   \maketitle
\section{Introduction}
Type III radio bursts are manifestations of transient energetic electron beams injected into the solar corona, propagating along the interplanetary magnetic field (IMF) lines \citep{ergun98, pick6, reid20}. As these beams traverse the corona, they trigger plasma waves, also known as Langmuir waves, which are then transformed into radio emission at the local plasma frequency or its harmonic components \citep{melrose17}. 
In the radio spectrograms, type III bursts are usually observed as intense emissions that drift in frequency over timescales of seconds--minutes and over a wide range of frequencies, from metric to decametric wavelengths \citep{wild50, lecacheux89, bonnin8}, making them detectable by ground-based instruments on Earth and various spacecraft within the heliosphere. 
The frequency of the radio emission is directly related to the plasma density, making type III bursts a valuable diagnostic tool for examining the inner heliosphere and the processes that drive solar active phenomena, such as solar flares and coronal mass ejections \citep{reid14, kontar17}.

The electron beams follow open magnetic field lines and can persist well beyond 1 astronomical unit (AU) (e.g., \citet{dulk85, boudjada20}), offering in-situ insights into the burst and ambient conditions of the heliosphere, including electron density, radio frequency drift, speed of the electron beams, and even potential direct detection of Langmuir waves (see \citet{gurnett76, gurnett77} and \citet{reid14} and references within). In addition, tracing the path of type III bursts provides a map of the density structure of the heliosphere, serving as a foundation for developing and testing density models.
Since radio observations below $\sim$10 MHz cannot be accomplished from the ground, it is important to combine high- and low-frequency observations from ground-based and space-borne instruments.
In this work, we perform a study of several type III radio bursts that occurred in close succession on April 3, 2019. We use remote observations of type III radio bursts detected by the Low-Frequency Array \citep[LOFAR]{lofar13} ground-based radio telescope and the Parker Solar Probe \citep[PSP]{fox16} spacecraft during Encounter 2 to study the sources of these radio emissions and to investigate the physical conditions responsible for their generation. Additionally, we incorporate results of two steady-state models of the solar corona: the Potential Field Source Surface (PFSS) model \citep{altschuler1969magnetic, schatten1969model} and the Magnetohydrodynamic Algorithm outside a Sphere (MAS) model \citep{mhd99}, to gain a better understanding of the coronal magnetic environment and its role in the acceleration of electrons. 
The ground-based LOFAR imaging observations provide valuable insight into the actual location of the burst sources. This research aims to expand upon current knowledge of the electron beams responsible for triggering type III radio bursts and the coronal conditions they experience. Gaining a deeper insight into this aspect is vital in comprehending other solar phenomena such as solar energetic particles and solar wind, and how they influence the near-earth space environment.

A number of recent studies investigate the physical mechanisms responsible for the generation of solar type III radio bursts. 
For example, \citet{chen13} investigated the association of type III bursts with flaring activities in February 2011, via combined multi-wavelength observation from the Solar Dynamic Observatory (SDO) instruments, as well as Wind/WAVE and ground-based instruments. They found that the SDO measurements indicated that type III emission was correlated with a hot plasma (7 MK) at the extreme ultraviolet (EUV) jet's footpoint. 
By using a triangulation method with the Wind and the twin STEREO spacecraft, \citet{bonnin8} reported the first measurements of the beaming characteristics for two type III bursts between 2007 -- 2008, assuming the source was located near the ecliptic plane (see also \citet{reiner9}). They concluded that the individual type III bursts have a broad beaming pattern that is roughly parallel to the Parker spiral magnetic field line at the source.
\citet{saint12} conducted a study on almost 10,000 type III bursts observed by the Nancay Radioheliograph between 1998 and 2008. Their analysis revealed discrepancies in the location of type III sources that may have been caused by a tilted magnetic field. Additionally, they found that the average energy released during type III bursts throughout a solar cycle could be comparable to the energy produced by non-thermal bremsstrahlung mechanisms in nano-flares.
\citet{morosan17} utilized LOFAR data to investigate the statistical characteristics of over 800 type III radio bursts within an 8-hour period on July 9, 2013. They discovered that the drift rates of type III bursts were twice that of type S bursts, and plasma emission was the primary emission mechanism for both types.

\citet{pulupa20} introduced a statistical overview of type III radio bursts during the first two PSP solar encounters. While the first encounter in November 2018 revealed a small number of bursts, the second encounter in April 2019 exhibited frequent type III bursts, including continuous occurrences during noise storms. They reported the characteristics of type III bursts with spectral and polarization analysis.

\citet{Krupar_2020} performed a statistical survey of 30 type III radio bursts detected by PSP during the second encounter in April 2019 and estimated their decay times, which were used to estimate the relative electron density fluctuations in the solar wind. They localized radio sources using a polarization-based-radio triangulation technique, which placed the sources near the modeled Parker spiral rooted in the active region AR12738 behind the plane of the sky as seen from Earth.

\citet{Cattell_2021} explored correlations between type III radio bursts and EUV emission in the solar corona. Using coordinated observations from PSP, SDO, and Nuclear Spectroscopic Telescope Array (NuSTAR) on April 12, 2019, they identified periodicities in EUV emission correlated with type III burst rates. The findings suggested impulsive events causing heating and cooling in the corona, possibly nano-flares, despite the absence of observable flares in X-ray and EUV data, which implies periodic non-thermal electron acceleration processes associated with small-scale impulsive events.

\citet{harra2021active} explored the origin of the type III radio bursts we are tackling in this paper and found that electron beams that triggered radio bursts may have emanated from the periphery of an active region that showed significant blue-shifted plasma.
More recently, \citet{badman22} observed a distinct type III radio burst using the PSP and LOFAR between 0.1 and 80 MHz on April 9, 2019, around 12:40 UT, six days after the occurrence of the event analyzed in our study. While no detectable flare activity was linked with the event, a type III noise storm was ongoing during the PSP encounter 2. The authors determined the type III trajectory and reconstructed its source using observations from Wind and STEREO spacecraft, as well as measuring related electron enhancement in situ.

In the last few years, we witnessed the emergence of modern instruments, such as LOFAR and PSP, that allowed to observe solar radio emissions with higher sensitivity from a better vantage point. Although type III bursts have been extensively studied \citep{dabrowski21}, there are still some unresolved issues regarding the exact mechanism of type III emissions.
For example, it is not yet clear how the electrons are accelerated to the high energies required to generate type III radio bursts, or what role the coronal magnetic field plays in this process.
Furthermore, there are inconsistencies between the observations and the models, which need to be resolved in order to gain a more complete understanding of the dynamics of the solar corona. Examples of these inconsistencies are the origin of the type III radio bursts and the discrepancy between the estimated plasma densities from the models and the observations. This paper aims to address these unresolved challenges by using new observations from LOFAR and PSP and models of the solar corona to study the physical mechanisms responsible for the generation of type III bursts.
The data analysis includes a combination of radio spectroscopy and imaging techniques to study the frequency, temporal and spatial variations of the radio bursts. 

The paper is organized as follows: In Section~\ref{s_obs}, we describe the observations of type III radio bursts made with LOFAR and PSP. In Section~\ref{s_methods} we explain the data analysis and modeling techniques used to study these events. In Section~\ref{s_results}, we present the results of our analysis, including an investigation of the potential physical mechanisms responsible for the generation of type III radio bursts, and a comparison of the observations with models of the solar corona. Finally, in Section~\ref{s_conclusions}, we summarize our findings and discuss their implications.
\section{Observations}
\label{s_obs}
A number of studies focused on observing the solar radio emissions during the second encounter of the PSP in late 2019 \citep{Krupar_2020, pulupa20, Cattell_2021, harra2021active, badman22}. In this study, our primary emphasis is directed towards investigating a set of type III radio bursts that took place on April 3, 2019, during the time interval spanning from $\sim$12:10 to 12:50 UT. This period coincided with the presence of two distinct active regions (ARs) on the Sun, denoted as AR12737 and AR12738. 
AR12737 was situated on the solar near side at coordinates E12$^o$N06$^o$. Notably, this region had 8 sunspots and exhibited a $\beta$ magnetic configuration according to the Hale magnetic classification \citep{hale_1919}. On the other hand, AR12738 was positioned on the solar far side at coordinates E140$^o$N02$^o$. Due to its remote location, detailed observations of the magnetic configuration and activity within AR12738 were unattainable during this time frame.

\begin{figure}
\centering
\includegraphics[width=0.9\hsize]{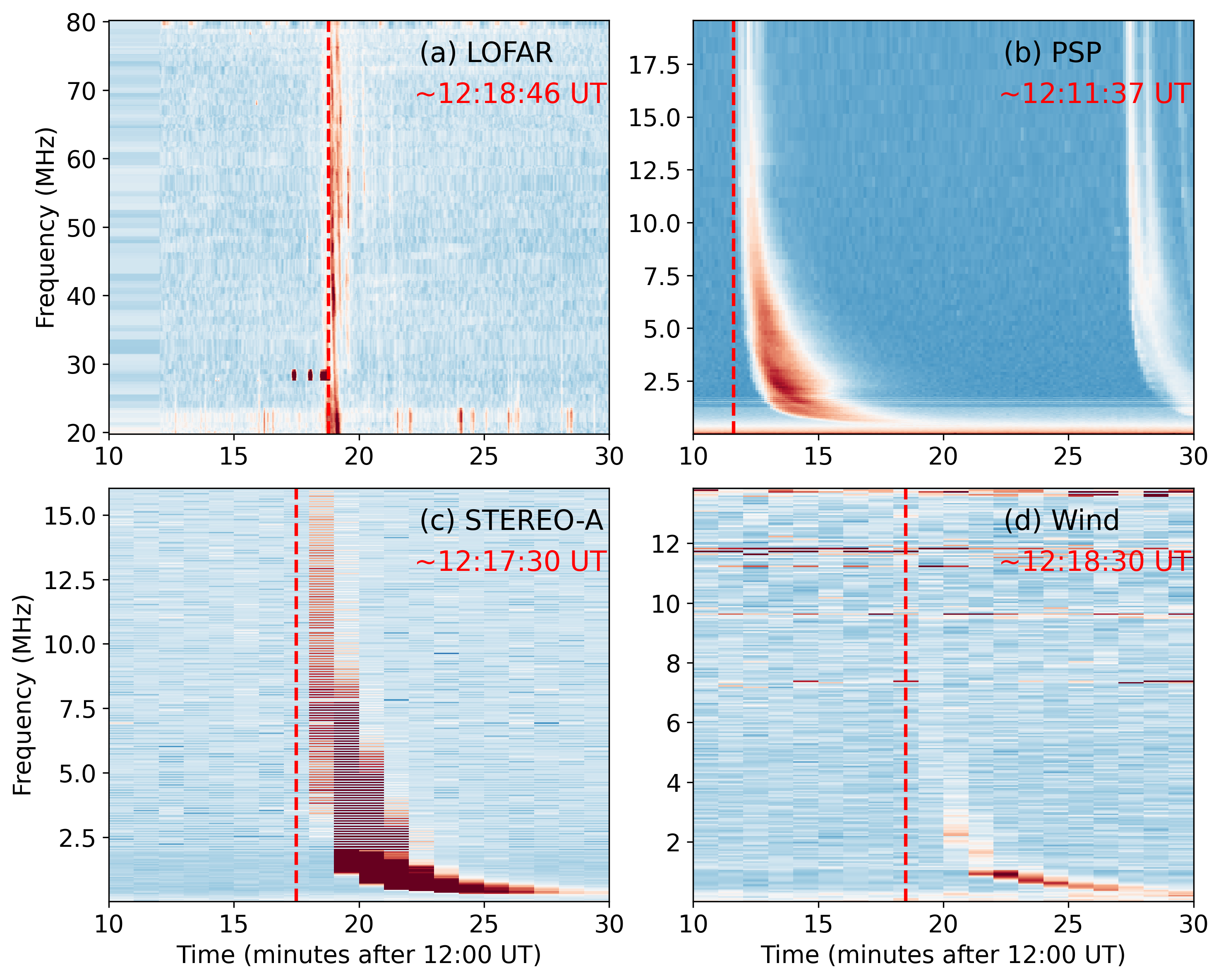}
\caption{Radio dynamic spectra for a single burst obtained from multiple instruments. The top-left panel is from the LOFAR/LBA instrument, the top-right is from the PSP/FIELDS instrument, the bottom-left is from the STEREO/SWAVES instrument, and the bottom-right is from the Wind/WAVES. The vertical red dashed line denotes the start time of the burst.}
\label{fig_alldyspec}
\end{figure}

We observed a group of intense type III radio bursts by four instruments (Wind/WAVES, PSP/FIELDS, STEREO-A/SWAVES, and LOFAR/LBA) while doing a regular survey. In Figure~\ref{fig_alldyspec}, we show the first type III burst within the time of this study as observed by the four instruments. By taking the 2$^{nd}$ derivative of the light curve at a specific frequency channels, we determine the start time of the burst which is denoted by the vertical red dashed line. The frequency bands used for obtaining the start time at each instrument are as follows: 6.97 MHz (Wind), 7.03 MHz (STEREO), 5.03 MHz (PSP), and 40.16 MHz (LOFAR).

\begin{figure}
\centering
\includegraphics[width=0.8\hsize]{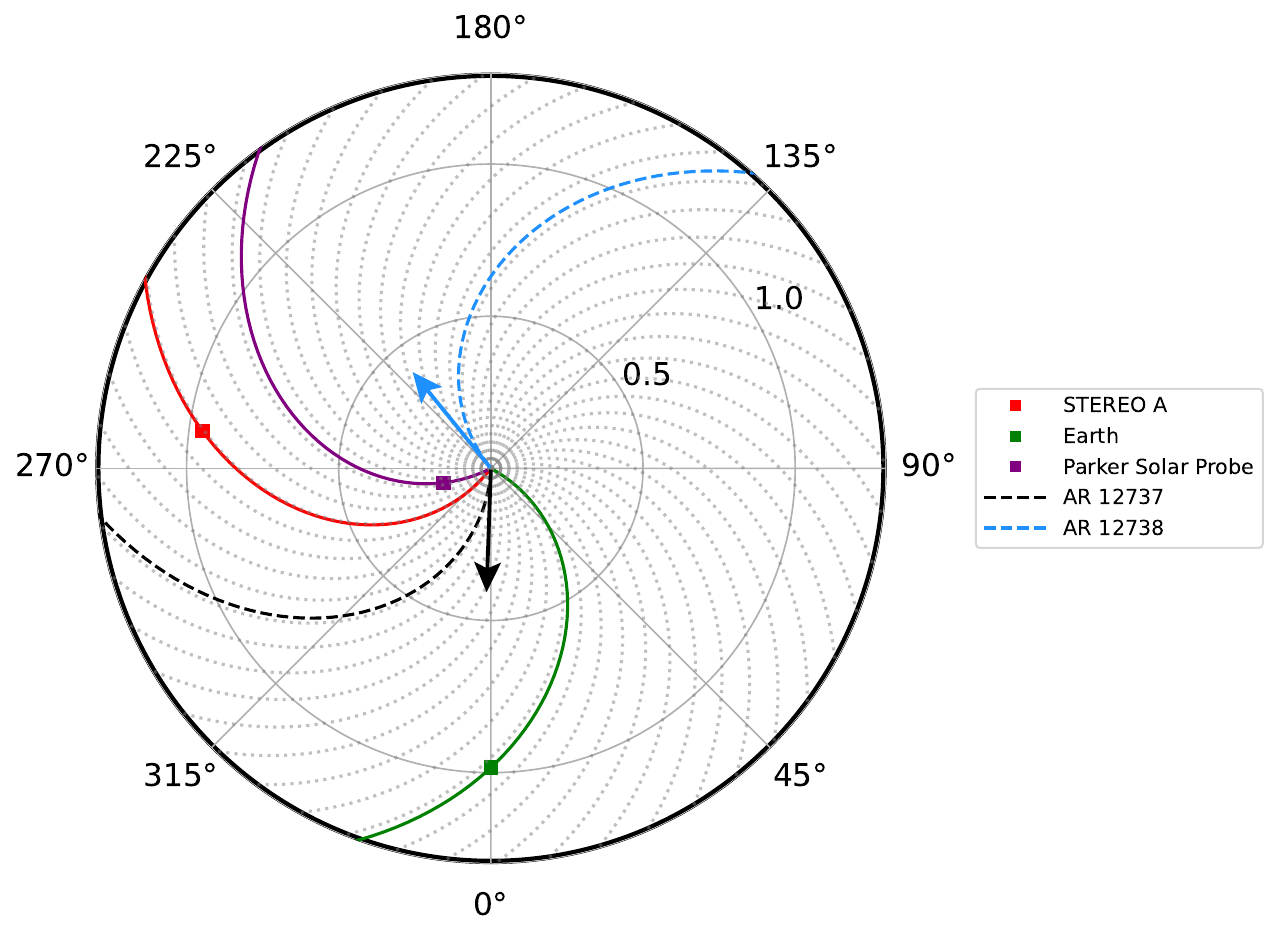}
  \caption{Top view of the spacecraft positions in the ecliptic plane at 12:15 UT on April 3, 2019, with the Sun-Earth line as the reference point for longitude. The earth's location is representative of the positions of LOFAR, Wind/WAVES, and GOES-15/XRS instruments. The spacecraft were connected back to the Sun by a 400 km/s reference Parker Spiral. The black arrow represents the longitude of AR12737, and the blue arrow represents the longitude of the AR12738. The gray dotted lines are the background Parker spiral field lines. The black dashed spiral shows the field line connected to the AR12737, and the blue dashed spiral is connected to the AR12738. The figure is generated using the Solar MAgnetic Connection Haus (\href{https://github.com/jgieseler/solarmach}{Solar-MACH}) tool \citep{Gieseler2023}.}
     \label{locations}
\end{figure}

\begin{figure*}[ht]
\centering
\includegraphics[width=\hsize]{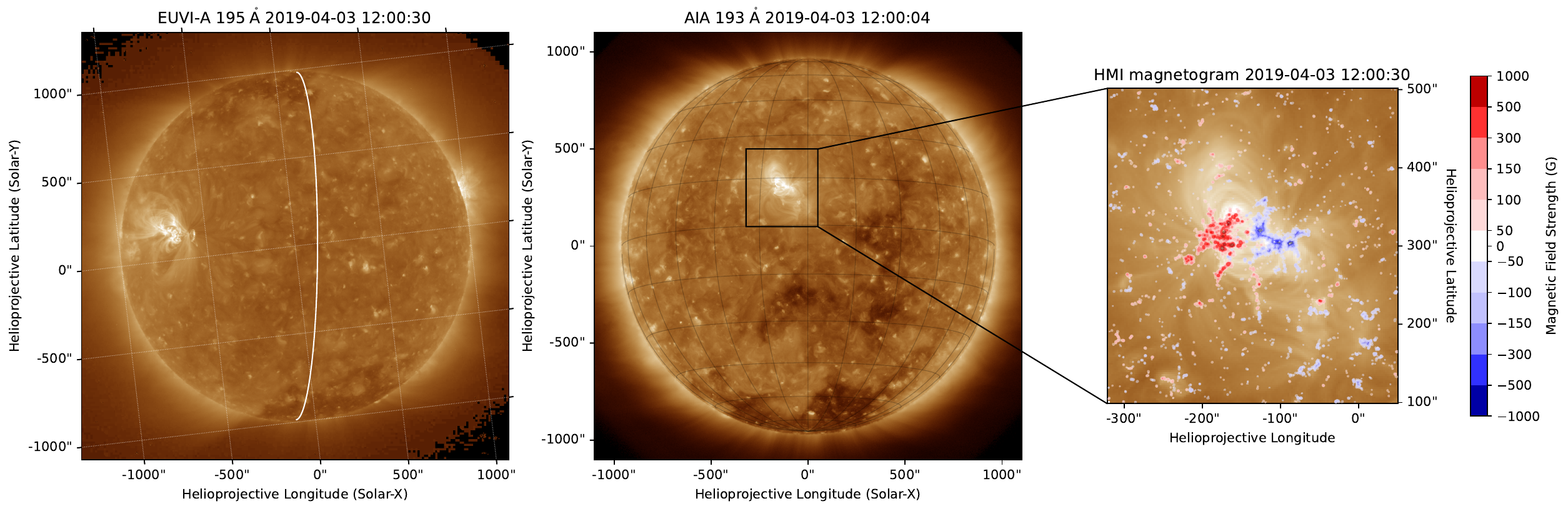}
\includegraphics[width=12cm]{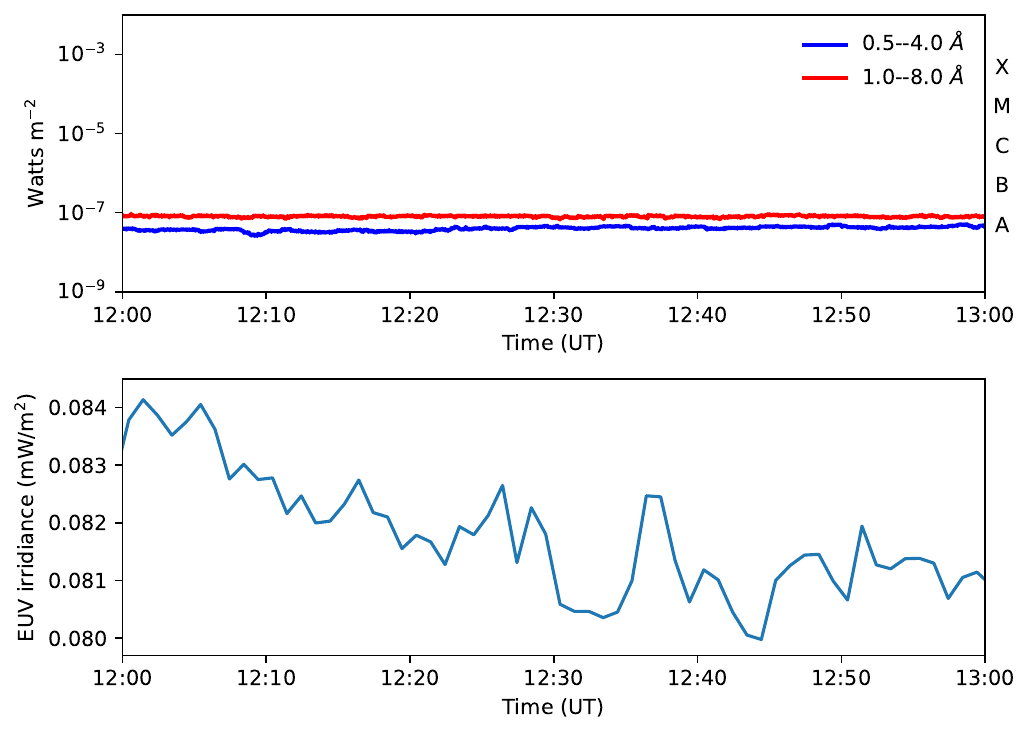}
\caption{Exploring the X-ray and extreme ultraviolet (EUV) emissions from the Sun. The top panel showcases a cutout region of the SDO/AIA 193$\AA$ image of the solar disk along with the STEREO-A EUVI 195$\AA$ point of view. The white curve is the limb of the solar disk as seen by AIA from the right side. The red-blue colors are the contours of the line-of-sight magnetogram from the SDO/HMI instrument. The levels are (50, 100, 150, 300, 500, 1000) Gauss. The middle panel shows the X-ray flux from the GOES-14 spacecraft shows minimum activity. The bottom panel shows the time series of the ESP Quad band from the SDO/EVE instrument, which shows the solar irradiance in the extreme ultraviolet (EUV) band.}
\label{soldisk_xrs}
\end{figure*}

We checked the relative orientations of the instruments with respect to Earth (Fig.~\ref{locations}). Since the PSP and STEREO spacecraft were almost aligned (close in an angular sense) with the Sun, the STEREO/EUVI image could be taken as what PSP would see (Fig.~\ref{soldisk_xrs}).
Figure~\ref{soldisk_xrs} shows how the solar disk looks like from the Earth perspective (using the SDO/AIA instrument) and from the eastern side where the PSP and STEREO were located at that time (using the STEREO/EUVI instrument).
The right panel shows a closer view of AR12737 with the contours of the photospheric magnetic field obtained from the Helioseismic and Magnetic Imager (HMI) onboard SDO.
From the GOES-15/XRS and SDO/EVE instruments in the panels below, they also confirm that there is no flaring activity at that time.

The solar disk was quiet, including the only one AR visible with no X-rays and no EUV transient emissions over this period.
Nevertheless, the very sensitive LOFAR telescope detected a number of bursts close to noon. We checked PSP data, and we found bursts there as well.
Meanwhile, from the EUVI and AIA images, we see that there are numerous small localized regions of relatively higher intensity, probably small-scale coronal brightenings spots or campfires (see \citet{young18, madjarska19, berghmans21}).
In the next subsections, we introduce the PSP and LOFAR instruments and their observations of the radio bursts.

\subsection{PSP Observations}
Parker Solar Probe (PSP) is a pioneering spacecraft with cutting-edge technologies, launched on August 12, 2018, to help resolving key questions about solar corona and solar wind \citep{fox16}.
To study the radio bursts, we use the level-2 data of the radio dynamic spectrum obtained from the FIELDS instrument suite \citep{bale16, pulupa17}, which can be downloaded from this website\footnote{PSP FIELDS data products: \url{http://research.ssl.berkeley.edu/data/psp/data/sci/fields/}}. The data file is in CDF format and the unit of the data values is converted from $V^2/Hz$ to $dB$ units using the formula
\begin{equation}
    I_{dB} = 10 \times log_{10}(I/10^{-16})
\end{equation}
The minimum power spectral density (PSD) of 10$^{-16}$ $V^2/Hz$ is used as a threshold for radio bursts according to \citet{pulupa20} for converting to decibels. Then, both the High-Frequency Receiver (HFR: 1.3 – 19.2 MHz) and the Low-Frequency Receive (LFR: 10.5 kHz – 1.7 MHz) data are combined into a single dynamic spectrum as shown in Figure~\ref{lofar_psp_burst_detect} with a full frequency range between 10.5 kHz - 19.2 MHz. The mean intensity value at each timestep over the full frequency range is subtracted from each frequency channel to clean the spectrum and minimize the noise level.
\subsection{LOFAR Observations}
The LOw Frequency ARray (LOFAR) radio telescope \citep{lofar13} is a powerful tool for studying the Sun at low radio frequencies ranging between 10 and 240 MHz. Its high sensitivity and high time resolution have enabled the detection of various solar phenomena, including radio bursts and CMEs, and the study of dynamic processes in the solar atmosphere on timescales of milliseconds.
The LOFAR dynamic spectrum from the beamformed radio observations is obtained by the Low-Band Antenna (LBA: 10 – 90 MHz) and can be downloaded from the LOFAR long-term archive (LTA)\footnote{LOFAR LTA: \url{https://lta.lofar.eu/}}. The High-Band Antenna (HBA: 110 – 190 MHz) data is not available for that time. For this day under study, the LOFAR data is available between 11:42 – 13:27 UT. To clean the spectrum, background subtraction is performed, which flattens the sensitivity (response) with the frequency of the LBA antennas. Basically, the mean spectrum along each frequency band is calculated and subtracted from the whole frequency band, the same applied to the PSP spectrum. This operation effectively removes the constant background from the spectrum. Then a Gaussian smoothing filter is applied to the spectrum using the \textbf{scipy.ndimage.gaussian\textunderscore filter} function with a sigma value of 1.5, which helps to reduce noise and variations in the data.
After that, the PSP and LOFAR spectra are combined together in a single plot within the same time interval. The bursts’ signals observed by the PSP occur earlier than those at LOFAR. This is due to the fact that the PSP spacecraft is much closer to the Sun and hence it detects the radio emissions earlier than LOFAR because of the shorter travel time of radio signals from the Sun. Therefore, the PSP dynamic spectrum must be shifted with respect to the LOFAR observations based on a calculation of the relative time travel of the radio emission from the Sun to PSP and to LOFAR. In addition, the time cadence of the PSP observations changes according to its distance from the Sun. On that day, the PSP data cadence was 7 seconds, while LOFAR’s is 1 second. Therefore, the LOFAR dynamic spectrum was downsampled to 7 seconds to match the time resolution of the PSP. Figure~\ref{lofar_psp_burst_detect} shows the resulting combined LOFAR-PSP spectrum on a logarithmic y-axis.
The LOFAR LBA frequency ranges between 19.82 -- 80.16 MHz and for the PSP is 10.55 kHz -- 19.17 MHz.

In order to detect the type III radio bursts automatically from the combined dynamic spectrum, we applied \citet{zhang18}'s algorithm which is based on the probabilistic Hough transformation that detects vertical bright edges in images, within a certain degree of deviation from the vertical direction.
\section{Methods}
\label{s_methods}
\subsection{Imaging of Radio Sources}
As part of our task, we developed an automated pipeline consisting of several modules that not only preprocessed and calibrated the LOFAR interferometric data to produce cleaned images of the Sun in the radio band \citep{zhang2022imaging}, but also utilized the resulting data to find the trajectory of the radio sources and sample the magnetic field and plasma parameters at their respective locations through modeling and simulation in subsequent modules.

\begin{figure}
   \centering
   \includegraphics[width=0.8\hsize]{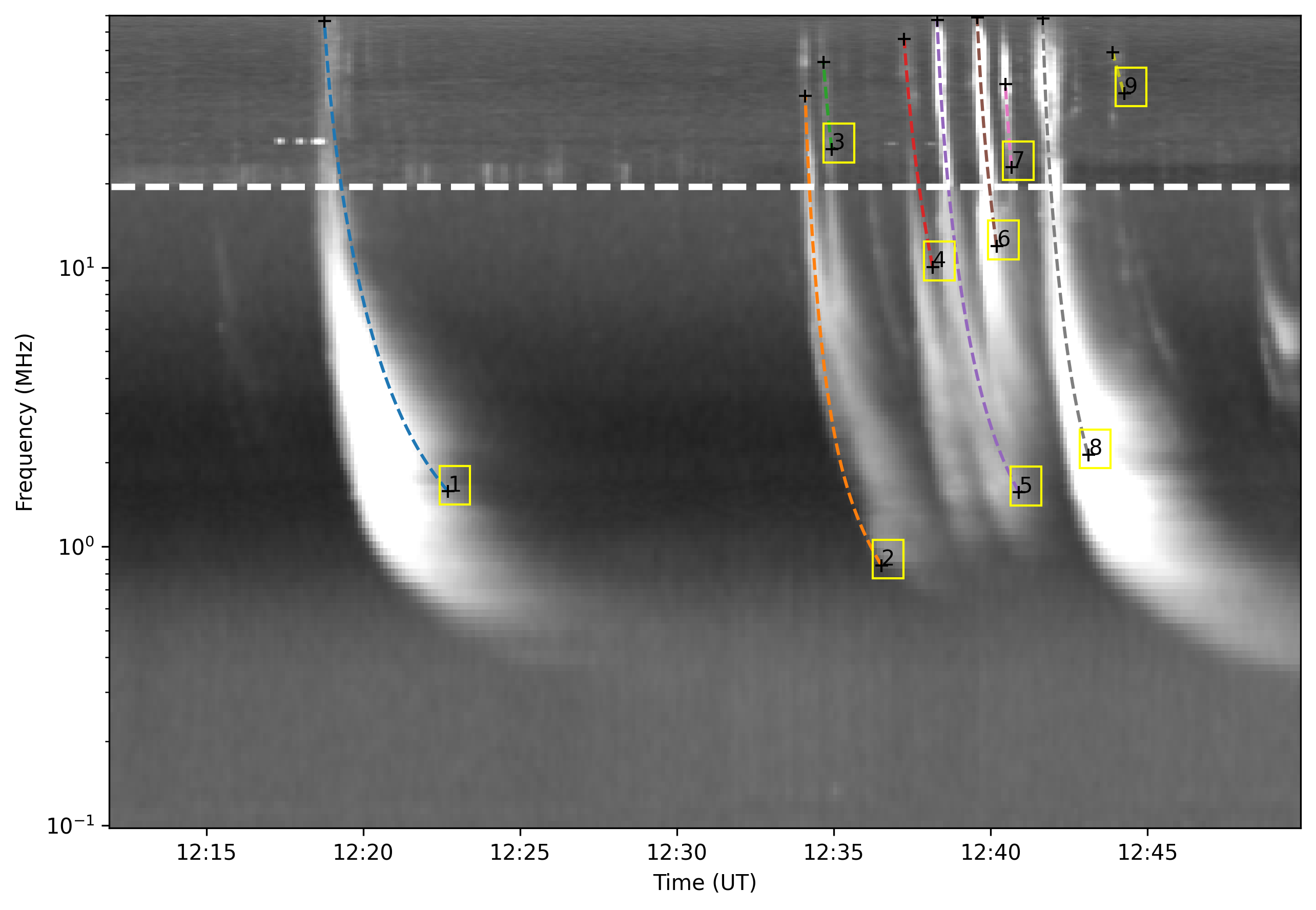}
      \caption{Automatic detection of type III radio bursts from the combined radio dynamic spectrum of LOFAR and PSP instruments. The dashed horizontal lines separates the LOFAR frequency range (top) and the PSP frequency range (bottom).}
    \label{lofar_psp_burst_detect}
   \end{figure}

First, we ran the burst detection algorithm \citep{zhang18} \footnote{Detection algorithm repository: \url{https://github.com/peijin94/type3detect}} on the combined dynamic radio spectrum of LOFAR and PSP (Fig.~\ref{lofar_psp_burst_detect}) in order to find the characteristics of each type III burst. We converted the spectrum into a binary map to isolate the bursts from the background. Then we applied the Hough transformation to get line segments of the features. For each type III burst, the line segments are grouped together into one group. To account for the interplanetary component within radio dynamic spectra, we employed the Parker electron-density model \citep{parker60} assuming a fundamental emission. This model enabled mapping between the time and frequency indices for each type III burst and subsequently converted electron densities into radial distances. Finally, a least-squares fitting method was applied to derive both the frequency drifts and the speed of the electron beams.

\begin{figure*}[ht]
\centering
\includegraphics[width=\hsize]{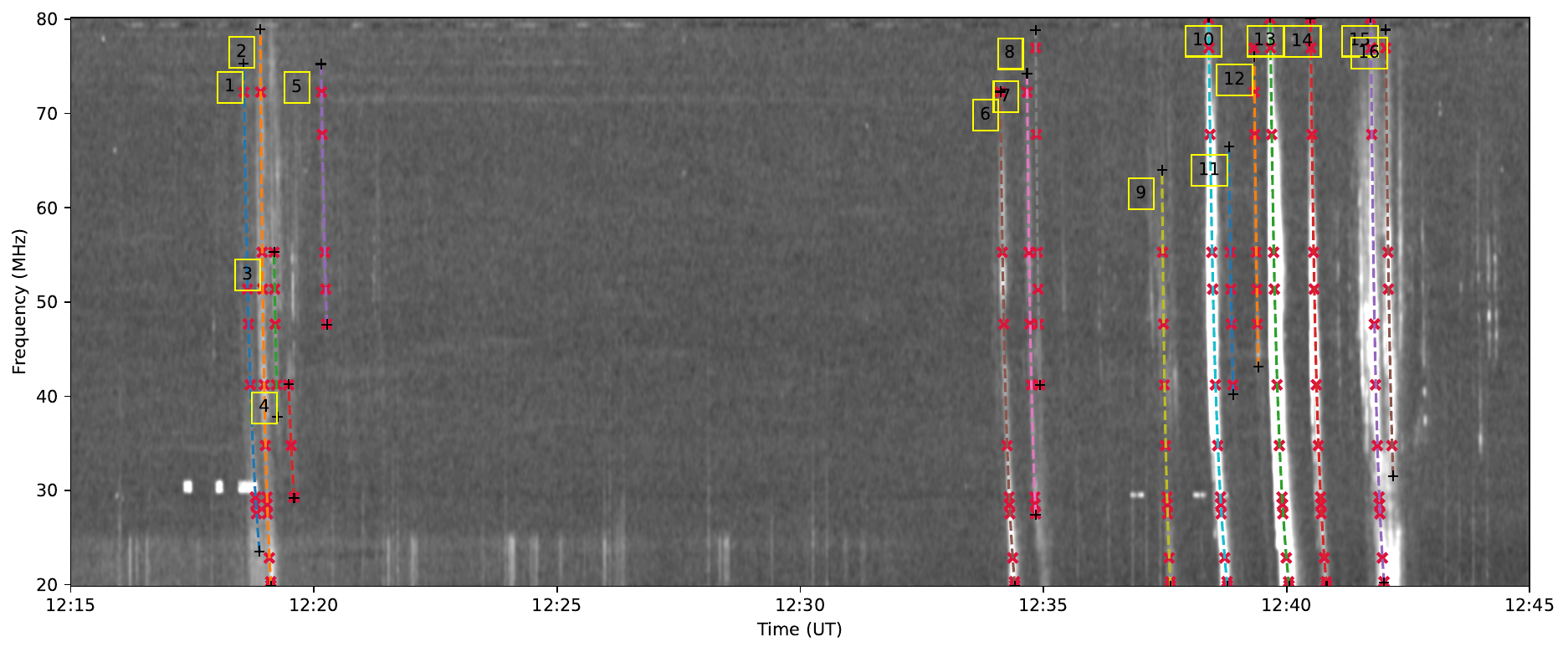}
\caption{Automatic detection of type III bursts observed by LOFAR. The red symbols along the fit lines are the $(f, t)$ coordinates of the image snapshots shown in Figure~\ref{fig_persistence}.}
\label{lofar_burst_detect}
\end{figure*}

After this step, we did the same for the LOFAR dynamic spectrum only (Fig.~\ref{lofar_burst_detect}) to find the $(f,t)$ pairs for every type III burst. Then we took snapshot frequencies for each burst defined by a list of 60 central frequencies between $\sim$20 -- 80 MHz from LOFAR LTA for the interferometric imaging. We obtained the interferometric data from LOFAR core and remote stations at the snapshot frequencies for all type III bursts. We used the concurrent observations of the radio source Tau-A in order to calibrate the interferometric observations. For that, we used the default preprocessing pipeline (DP3) \citep{dppp18} for preliminary processing and calibrating the measurement sets (MS). Finally we obtained the cleaned images of the radio sources by using $w$-stacking clean (WSClean) algorithm \citep{wsclean14} at only the time indices in the MS files that are equivalent to the snapshot frequencies.
\begin{table*}[h!]
\centering
\caption{Characteristics of the type III bursts detected via the automatic algorithm from the combined spectrum.}
\label{table_bursts}
\resizebox{0.9\textwidth}{!}{%
\begin{tabular}{ccccccc}
\hline
\begin{tabular}[c]{@{}c@{}}Burst\\ ID\end{tabular} & \begin{tabular}[c]{@{}c@{}}Start Time\\ (UT)\end{tabular} & \begin{tabular}[c]{@{}c@{}}End Time\\ (UT)\end{tabular} & \begin{tabular}[c]{@{}c@{}}Start Frequency\\ (MHz)\end{tabular} & \begin{tabular}[c]{@{}c@{}}End Frequency\\ (MHz)\end{tabular} & \begin{tabular}[c]{@{}c@{}}Frequency Drift\\ (MHz s$^{-1}$)\end{tabular} & \begin{tabular}[c]{@{}c@{}}Beam Speed\\ (c)\end{tabular} \\ \hline
1                                                  & 12:18:45                                                  & 12:22:42                                                & 76.44                                                           & 1.57                                                          & 0.892                                                                   & 0.044                                                           \\
2                                                  & 12:34:05                                                  & 12:36:31                                                & 41.24                                                           & 0.86                                                          & 0.241                                                                   & 0.119                                                           \\
3                                                  & 12:34:40                                                  & 12:34:56                                                & 54.44                                                           & 26.54                                                         & 3.992                                                                   & 0.046                                                           \\
4                                                  & 12:37:14                                                  & 12:38:09                                                & 66.03                                                           & 10.02                                                         & 4.006                                                                   & 0.046                                                           \\
5                                                  & 12:38:17                                                  & 12:40:54                                                & 76.92                                                           & 1.57                                                          & 0.77                                                                    & 0.066                                                           \\
6                                                  & 12:39:34                                                  & 12:40:11                                                & 78.86                                                           & 11.93                                                         & 3.192                                                                   & 0.062                                                           \\
7                                                  & 12:40:28                                                  & 12:40:40                                                & 45.34                                                           & 22.9                                                          & 3.21                                                                    & 0.067                                                           \\
8                                                  & 12:41:39                                                  & 12:43:06                                                & 78.21                                                           & 2.13                                                          & 1.555                                                                   & 0.093                                                           \\
9                                                  & 12:43:53                                                  & 12:44:15                                                & 59.07                                                           & 42.13                                                         & 2.424                                                                   & 0.013                                                           \\ \hline
\end{tabular}%
}
\end{table*}

\begin{figure*}[ht]
\centering
\includegraphics[width=\hsize]{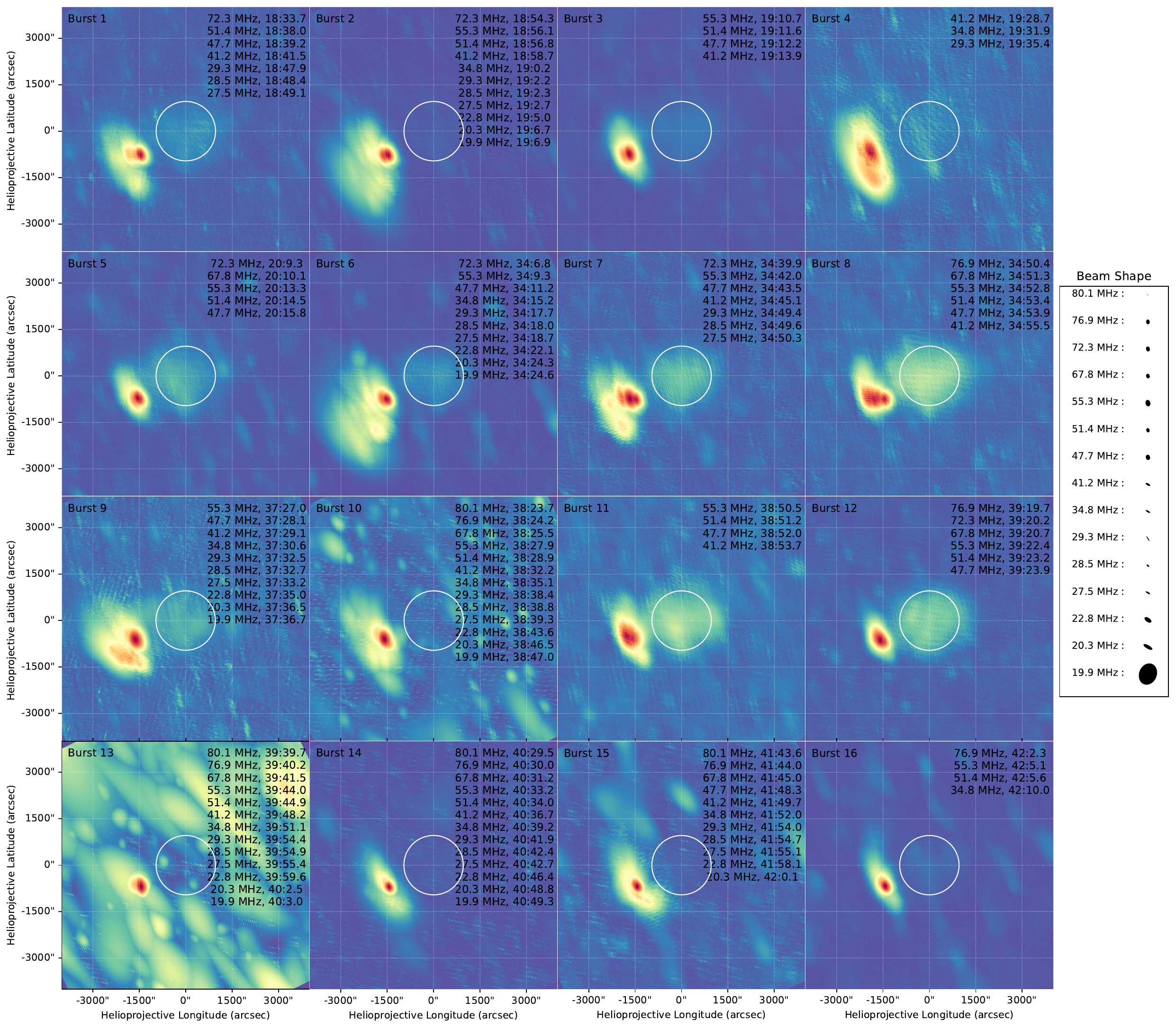}
\caption{Persistence imaging for the 16 type III bursts detected in the LOFAR dynamic spectrum. The label shows the observation frequencies in MHz and times in (minutes:seconds from 12:00:00 UT). Here, the color coding is not absolute, but rather each panel has its own color code.}
\label{fig_persistence}
\end{figure*}

After processing and cleaning the interferometric measurements of LOFAR, we explored the observations of each burst individually.
Out of the 60 frequency bands in the LOFAR LTA, we chose 54 frequency bands that have unique integer numeric, between 19.92 - 80.08 MHz.
For each burst, at each timestamp, the nearest frequency of the fit model to the list of chosen frequencies is picked as the snapshot frequency at that particular timestamp.
This process was repeated for all the 16 type III bursts detected in the LOFAR dynamic spectrum in order to obtain snapshot images for each type III burst (Fig.~\ref{fig_persistence}).
For each type III burst, we applied persistence imaging in order to create a continuous display of the radio emissions \citep{thompson2016persistence}.

Persistence imaging enables the creation of a clearer and more informative image.
In the context of a time-ordered series of images, a method of persisting pixel values can be employed as follows: for each image, compare the value of each pixel to its corresponding value in the previous persistence image in the series. If the pixel value in the current image is brighter than its corresponding pixel in the previous image, replace the previous value with the current one; otherwise, retain the previous value. This process generates a new image, referred to as the current persistence image, which serves as the basis for the subsequent evaluation of the next image in the series. This evaluation involves a pixel-by-pixel comparison between the current image and its associated persistence image, allowing for the identification of any changes or patterns that may have occurred over time. The mathematical background is explained in Appendix~\ref{append_a}.

In order to estimate the locations of the type III sources in 3D space, we combined observations with modeling. We used magnetogram data from the Global Oscillation Network Group project (GONG) \citep{gong96}. We constructed a grid of footpoints on the GONG map over two longitudinal belts around the two active regions AR12737 and AR12738, which are the two potential candidates source regions for the group of type III bursts under study. These points are used as the seed points for tracing the coronal magnetic field lines using pfsspy python package\footnote{Pfsspy tool: \url{https://pfsspy.readthedocs.io/}}, which is a robust implementation in python of the PFSS model developed by \citet{pfss20}.
Using the major and minor axes of the beam size, we estimated the radius of the radio source using Equation~\ref{source_radius}, which was used to approximate the source size.
Since we already obtained the $(x,y)$ positions of the type III sources in the plane of the sky (POS) through LOFAR observations, now it is necessary to determine their corresponding $z$ position to have an overall understanding of their spatial distribution. Therefore, we employed \citet{badman22}'s approach here, assuming that the type III bursts were from harmonic emission. First, we found the radial distance from the Sun in POS ($r_{pos}$) of the radio source on the POS (Eq.~\ref{rpos}). Second, we calculated the sources' radial distance ($r_{model}$) using the 2.5$\times$Newkirk electron-density model \citep{newkirk61, newkirk67}. The 2.5 fold factor is taken to incorporate the effects of scattering and overdensity (streamers) beyond the nominal Newkirk quiet Sun model. The MAS model results (Fig. \ref{forward2d}) show streamers above the eastern limb, supporting the inclusion of such a factor.
Lastly, we estimated the $z$ location of the type III sources (Eq.~\ref{zpos}).
We proceeded with the $+z$ solution because the theory precludes emission behind POS in this region of high-density gradients (i.e., the emission would be absorbed by passing through the high-density regions of the corona). More details are explained in Appendix~\ref{append_b}.
\begin{equation}
     r_{source} = \sqrt{(b_{major}^2 + b_{minor}^2)}
    \label{source_radius}
\end{equation}
\begin{equation}
    r_{pos} = \sqrt{(x^2 + y^2)}
    \label{rpos}
\end{equation}
\begin{equation}
     z = \sqrt{(r_{model}^2 - r_{pos}^2)}
    \label{zpos}
\end{equation}
The result of the deprojection of the type III sources for the 6$^{th}$ burst are shown in Figure~\ref{fig_3figs} with 70\%-contours made for 10 frequencies on the extrapolated magnetic field lines. The red dashed line is a spline fitting curve that represents the trajectory of the centroids of the radio sources. The black arrow points towards the Earth's line of sight (LOS). It is worth to mention that the axes direction in the POS of LOFAR images are different in the 3D space. The $(x,y)$ coordinates in the POS are translated into $(y,z)$ in the 3D space, and $z$ in the POS is translated into $x$ in the 3D space.
\subsection{Modeling}
To explore the characteristics of the coronal plasma environment during the studied events, we used Predictive Science Inc. (PSI)'s standard coronal solutions from magnetohydrodynamic (MHD) simulations originating from the Magnetohydrodynamic Algorithm outside a Sphere (MAS) code \citep{mhd99}. The data is available on the PSI's data archive\footnote{Predictive Science Inc.: \url{https://www.predsci.com/mhdweb/home.php}}. We obtained the PSI MAS coronal solution (a thermodynamic-with-heating MHD model) on April 3, 2019, at 12:00 UT with a simulation result ID of \texttt{hmi\_\_med-cor-thermo2-std01\_\_med-hel-poly-std01}.
Initially, we calculated the angle between the burst's source radial vector and the line of sight (LOS). Moreover, we calculated the complement angle, which is the separation angle between the burst's radial vector and the plane of the sky (POS) from the Earth's perspective.
Subsequently, we utilized the complement angle to derive the Carrington longitude \citep{Thompson2006}, facilitating the extraction of a longitudinal segment from the MAS datacube, as if it were in the POS. Following this, the selected data slice was fed into the FORWARD model—a toolset responsible for generating synthetic coronal maps of observable quantities describing the plasma state.
For extracting the longitudinal slices from the MAS data, we utilized the psipy python package\footnote{Psipy repository: \url{https://github.com/predsci/PsiPy}}.
The MAS datacube is specifically defined on a spherical grid and represents a steady-state MHD model. Owing to the inherent attributes of this datacube, the utilization of the FORWARD toolset proves more practical and advantageous for our objective.
In Figure~\ref{forward2d} we show the first radio contour of the 6$^{th}$ type III burst on top of the equivalent 2D maps for 6 plasma parameters, as an example. The plasma parameters are, from left to right and from the top to bottom: plasma density, plasma temperature, magnetic field strength, plasma beta parameter, the total plasma pressure, and the Alfven speed, By taking the value of these physical plasma quantities at the centroids' coordinates of the type III sources at each frequency band, we obtained estimates of local plasma conditions shown in Figure~\ref{scatterplot} for the 6$^{th}$ type III burst, as an example.
\section{Results and Discussion}
\label{s_results}
\subsection{Detection and characterization of type III radio bursts}
We found that the radio waves arrived at STEREO one minute before they arrived at Wind (Fig.~\ref{fig_alldyspec}). However, the difference between the $+z$ and $-z$ positions of the burst this close to the Sun in terms of light travel time is $\sim$10 seconds ($\sim$4 $R_\odot$), which is within the time resolution of the observations (1-min time resolution). Thus, we cannot confidently conclude whether the emission arrived at one spacecraft first and the other second.

Figure~\ref{lofar_psp_burst_detect} shows the combined dynamic spectrum from both LOFAR and PSP. The free parameters of the auto-detection algorithm do not have the same values as for detection the type III bursts in the LOFAR spectrum alone.
Upon visual examination, we observed that the detection algorithm effectively identified type III bursts in the LOFAR dynamic spectrum (Fig.~\ref{lofar_burst_detect}), but it had limitations in detecting type III bursts in the combined spectrum of the LOFAR and PSP, as well as missing segments of the detected bursts and a few bursts entirely.
This could be due to the increased frequency drift and dispersion of the radio bursts at lower frequencies, which made it a challenging task for the detection algorithm.
We captured 9 type III bursts from the combined dynamic spectrum and their characteristics are reported in Table~\ref{table_bursts}.
However, the detection algorithm performed better on the LOFAR dynamic spectrum only and we traced 16 type III bursts.
\subsection{Imaging of radio emission sources}
Figure~\ref{fig_persistence} shows the persistence imaging for the 16 type III bursts in the LOFAR dynamic spectrum (Fig.~\ref{lofar_burst_detect}). The observation frequencies and timestamps of the snapshot images used to produce the persistence image are shown at the top-right corner of each image.
From visual inspection of Figure~\ref{fig_persistence}, it seems that all the type III emissions originated from the same quadrant in the images (south-east direction on the solar disk), although there was no active region presented at that location except for a single active region nearby the central meridian (Fig.~\ref{soldisk_xrs}).
Based on the imaging data presented in Figure~\ref{fig_persistence}, we chose one representative type III burst (No. 6) for a single-burst analysis in this paper, as it shares similarities in extent and location with other bursts.
To determine the spatial connection between the sources of radio emissions and the coronal magnetic field, a three-dimensional projection of the radio source contours onto the extrapolated coronal magnetic field via the PFSS model was employed (Fig.~\ref{fig_3figs}). The result indicates a discernible south-eastward propagation of the radio sources relative to the Earth's perspective, with no open field line crossing the radio sources. In Figure~\ref{fig_3figs}, we performed an extrapolation only over the two active regions presented on the solar surface at that time. However, when we extrapolated the magnetic field over the entire solar surface, we noticed that the radio sources are aligned with the lower part of large-scale closed field lines, and are placed onto the open field lines emanating from the southern coronal hole. No open field lines crossing the radio sources are observed.

We note that the PFSS modeling is limited by the fact that AR12738 is behind the limb on April 3 as observed from Earth. Consequently, the magnetic data available to us could be around two weeks old or more. This might limit the reliability of PFSS extrapolation for that region during that specific timeframe.

From Figure~\ref{fig_3figs}, the results suggest several potential origins of these type III radio emissions:
\begin{itemize}
    \item they could be triggered in a closed-field lines structure such as large-scale coronal loops, given that the radio sources are aligned to closed-field lines geometry in the southern hemisphere;
    \item they could be triggered by electron beams that are accelerated from an open-field active region \citep{kong2018observational}. However, from the PFSS model, we found no evidence for magnetic connectivity from both ARs on the Sun at that time;
    \item they may result from electron beams that are accelerated in the corona due to expanding magnetic fields from plasma upflows in the active region \citep{del2011single, harra2021active}.
\end{itemize}
Our findings indicate a notable inverse relationship between imaging quality and the level of solar radio emission brightness (e.g., for type III bursts No. 10 and 13, for instance). This observation is due to the leakage of solar radio emission into the side lobes of the calibrator beam, which disrupts the accuracy of calibration solutions.
\subsection{Plasma diagnostics and magnetic field analysis}
\begin{figure*}[ht]
    \centering
    \begin{subfigure}[b]{0.45\linewidth}
        \centering
        \includegraphics[width=\linewidth]{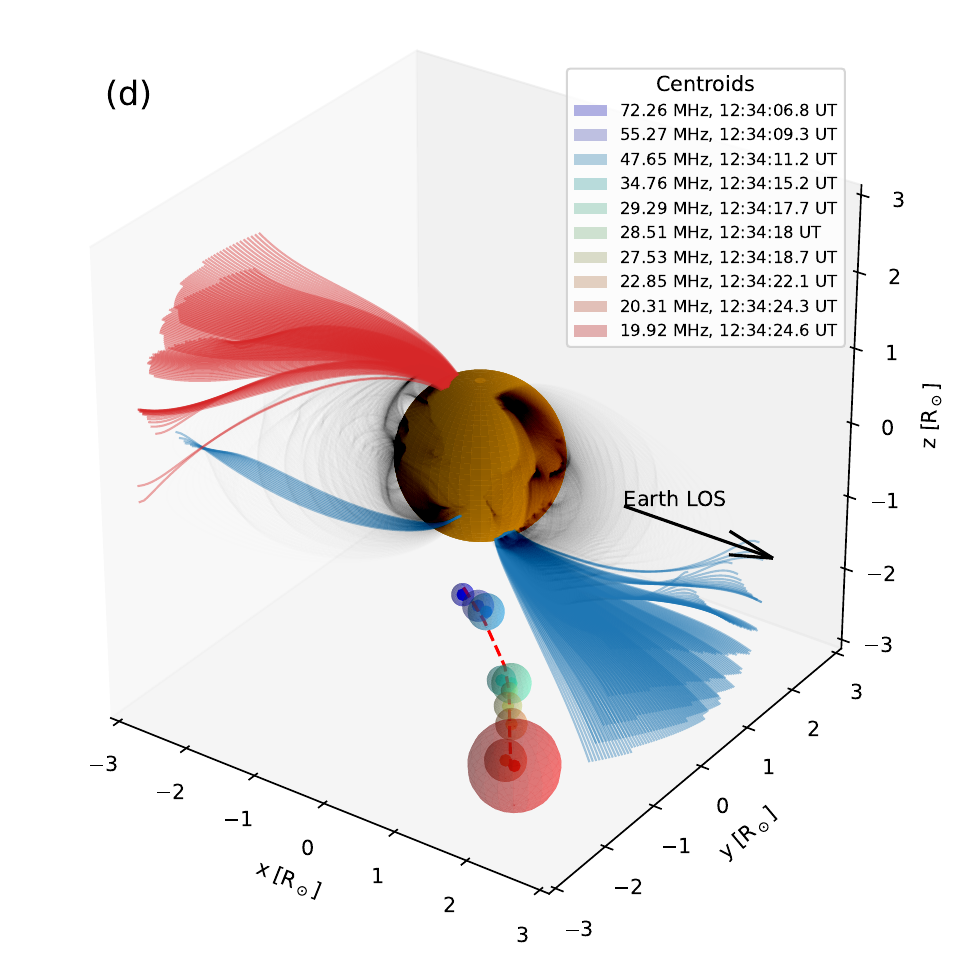}
        \label{subfig:a}
    \end{subfigure}
    \hfill
    \begin{subfigure}[b]{0.45\linewidth}
        \centering
        \includegraphics[width=\linewidth]{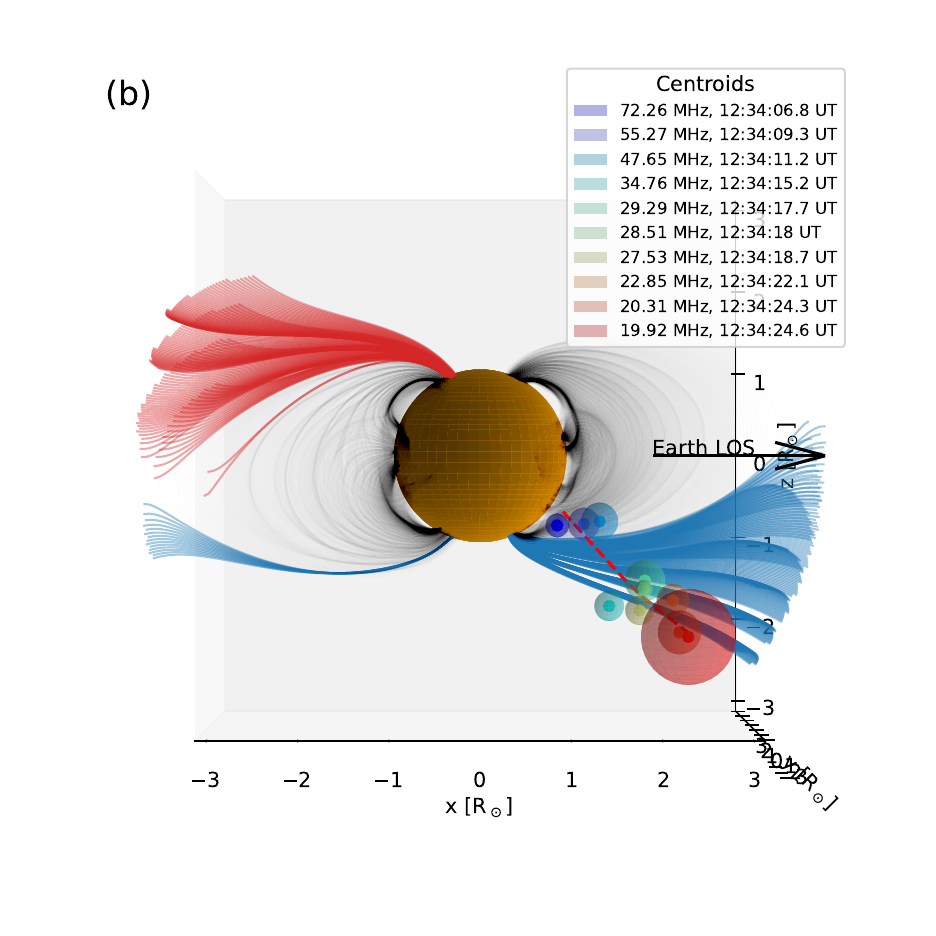}
        \label{subfig:b}
    \end{subfigure}
    
    \begin{subfigure}[b]{0.45\linewidth}
        \centering
        \includegraphics[width=\linewidth]{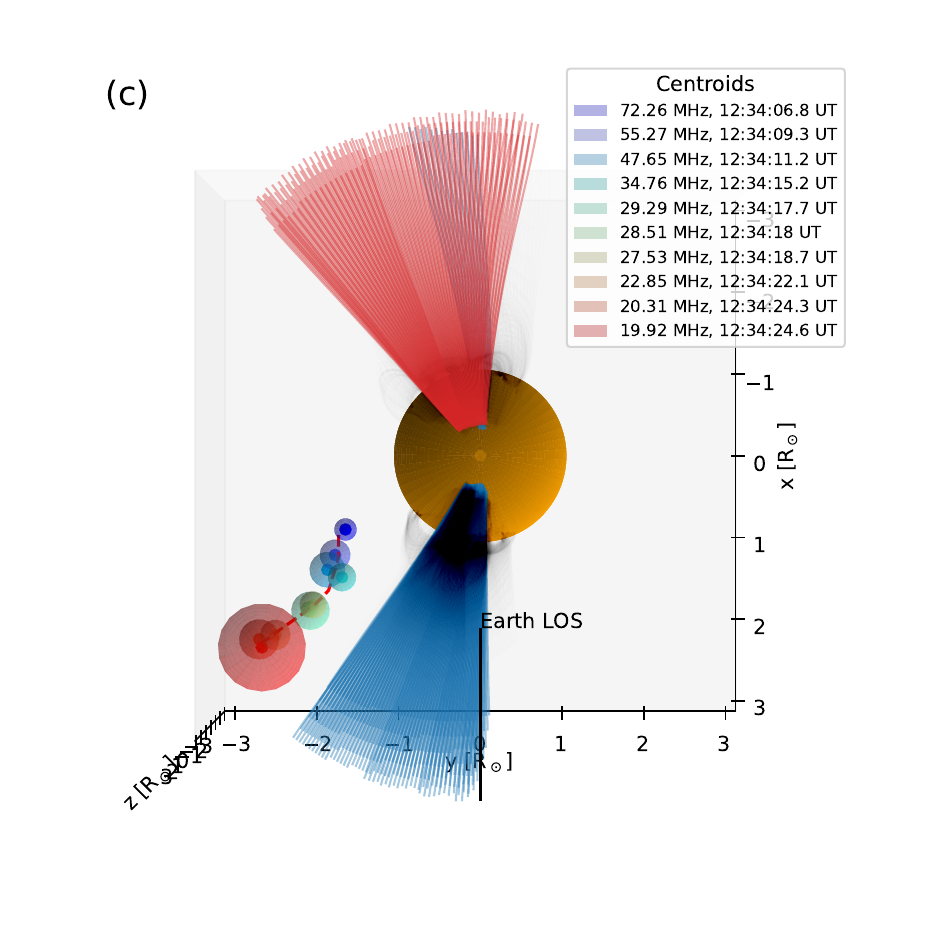}
        \label{subfig:c}
    \end{subfigure}
    \hfill
    \begin{subfigure}[b]{0.45\linewidth}
        \centering
        \includegraphics[width=\linewidth]{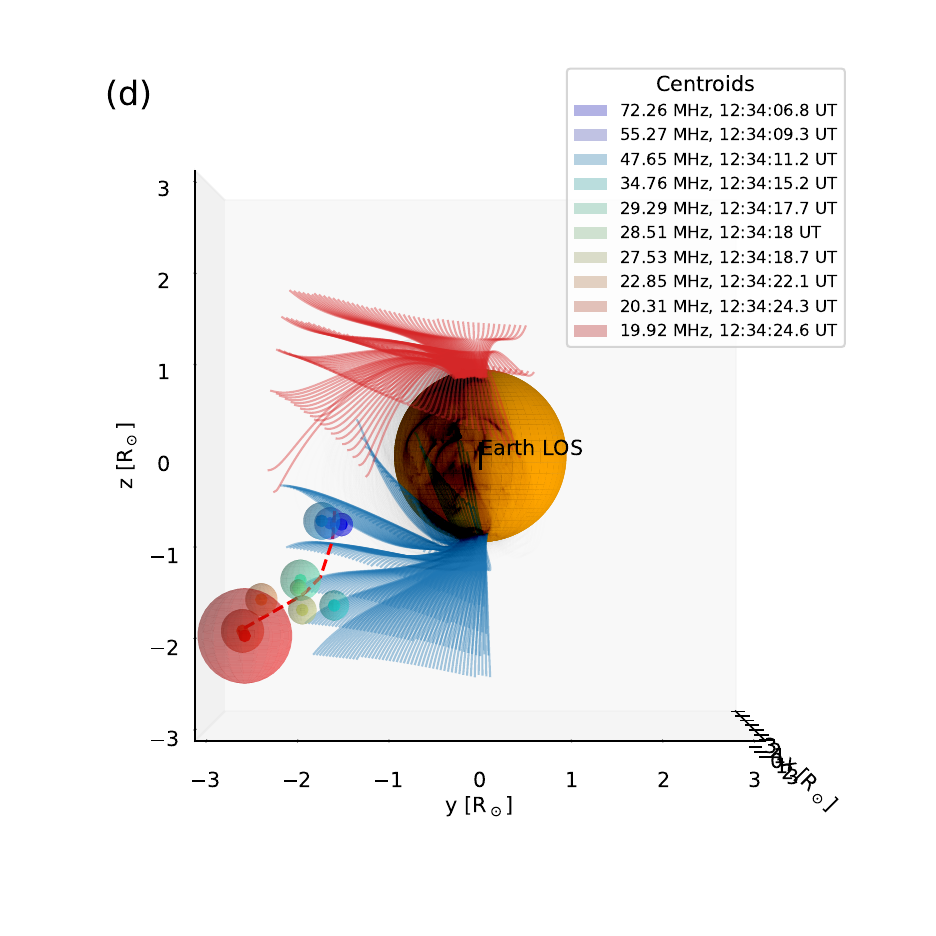}
        \label{subfig:d}
    \end{subfigure}
    \caption{Different viewing angles for the de-projection of the radio sources of the 6$^{th}$ burst using the 2.5$\times$Newkirk electron-density model on the PFSS solution. The black arrow points toward the Earth LOS. The $yz$ plane is the plane of sky as seen from the Earth. The red dashed line is a spline curve fit for the sources' centroids. The red, black, and blue curves are open northern, closed, and open southern field lines, respectively. The opacity of the closed field lines is decreased for a better visualization.}
    \label{fig_3figs}
\end{figure*}

\begin{figure*}[ht]
\centering
\includegraphics[width=0.98\textwidth]{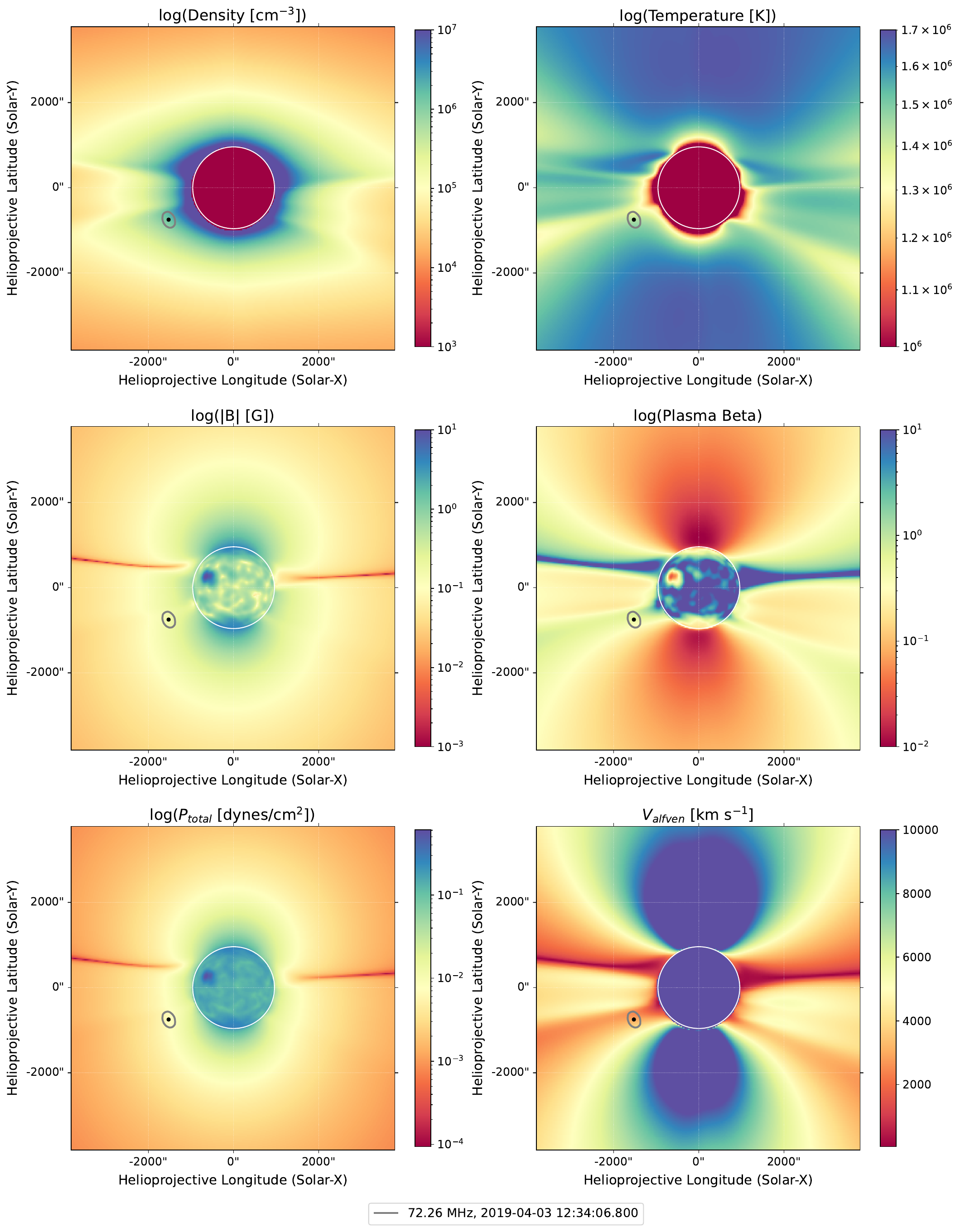}
\caption{Synthesized maps of plasma parameters obtained using the FORWARD toolset, with the 70\%-contour of radio emission of the 6$^{th}$ burst at the first timestamp (12:34:06.8 UT) at the frequency of 72.26 MHz depicted on top of the 2D plane-of-sky cuts. The left column represents, from top to bottom, plasma density, magnetic field, and the total plasma dynamic pressure. The right column represents, from top to bottom, the temperature, plasma beta, and the Alfven speed.}
\label{forward2d}
\end{figure*}

Considering the observed alignment of radio sources in Figure~\ref{fig_3figs} and the case depicted in Figure~\ref{forward2d}, it becomes evident that radio sources at higher frequencies (indicating proximity to the Sun) align with a streamer-like structure near the equator within the coronal model. This structure is characterized by elevated plasma beta, reduced coronal temperature, and diminished Alfven speed.
The coronal plasma density was relatively homogeneous with no prominent structures, probably due to the model resolution.

\begin{figure*}[ht]
\centering
\includegraphics[width=\textwidth]{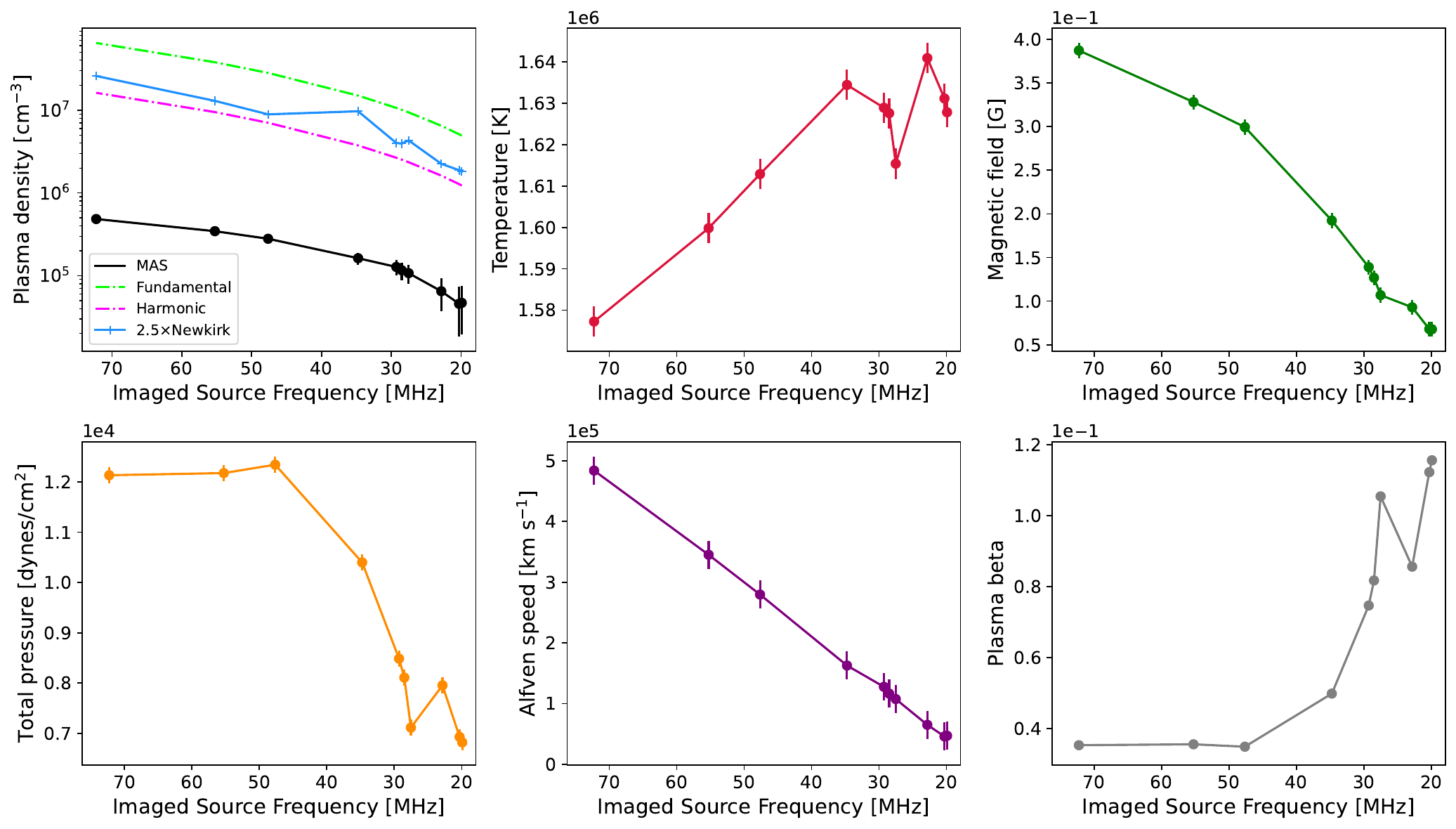}
\caption{Coronal plasma parameters sampled from the 2D maps by the source centroids. The top panel shows, from left to right, plasma density profiles from the MAS model, 2.5×Newkirk model, and the theoretical densities under the fundamental and harmonic assumptions, plasma temperature, and magnetic field. The bottom panel shows, from left to right, the total plasma dynamic pressure, Alfven speed, and plasma beta. The x-axis is inverted to to demonstrate a progression of increasing radial distance from the Sun as one moves towards the right.}
\label{scatterplot}
\end{figure*}

The location of radio sources of all the bursts were in the same quadrant as seen from Earth. Therefore, we assumed that the former description applies for all bursts.
We also found that the radio sources were confined between the equatorial sheet and the southern coronal hole and moving along that boundary.
Figure~\ref{scatterplot} shows the variability of the coronal plasma quantities at the radio sources' centroids, taken from FORWARD maps in Figures~\ref{forward2d}, at different frequencies for the 6$^{th}$ burst.
To estimate the error bars, we initialized random centroids, within the limits of the 70\%-contours of the radio emissions, to sample the plasma quantities at those locations. Then the standard error (SE) is calculated using Equation~\ref{error}, where $\sigma$ is the standard deviation, and $n$ is the number of points.
\begin{equation}
    SE = \frac{\sigma}{\sqrt{n}}
\label{error}
\end{equation}
The coronal temperature was increasing with radial distance, which implies there may have been some heating locally.
The behavior of the coronal magnetic field, the plasma total dynamic pressure, and the Alfven speed were decreasing over distance as expected.
Finally the value of plasma beta parameter started increasing sharply around 40 MHz, which implies that the plasma pressure became more dominant than the magnetic pressure around that distance from the Sun (for a 2.5$\times$Newkirk model, it is 1.89 $R_\odot$ assuming a fundamental emission, or 2.57 $R_\odot$ assuming a harmonic emission).

The top-left panel of Figure~\ref{scatterplot} shows a comparison between the density profiles of the MAS model, the 2.5$\times$Newkirk model, and the theoretical expected density profiles under the fundamental and harmonic assumptions.
Although the Newkirk density model provided a useful approximation for determining the height of radio sources in the corona, it is not entirely accurate due to a number of its underlying assumptions, for instance, the assumption of a steady state and the spherical symmetry of the corona, which do not always apply.
Therefore, we tried to use the MAS density values to estimate the depth along the LOS of the radio source since it is supposed to give a more realistic result.

We found that the plasma density obtained from the MAS and FORWARD modeling results were significantly lower compared with the 2.5$\times$Newkirk density model and the theoretical expected density obtained from the classical relation in Equation~\ref{plasma_freq}, where $f_p$ is the plasma frequency (in MHz) and $n_e$ is the electron density (in cm$^{-3}$).
\begin{equation}
    n_e = \left(\frac{f_p}{8.98 \times 10^{-3}}\right)^2
\label{plasma_freq}
\end{equation}
The required density from the fitted Newkirk model is much higher ($\sim$10 times) than what is obtained from the MAS model, even after accounting for the 2.5$\times$ enhancement already applied to the standard Newkirk model. This implies the discrepancy cannot be fully explained by the density enhancement factor alone. Furthermore, the imaging places the radio sources near a streamer which is an overdense region in the MAS model, so it seems unlikely the source's apparent location in the model is wrongly attached to a less dense feature, as there are not denser options available.
The apparent source positions from the imaging are likely too high, possibly due to scattering effects \citep{kontar_2019, kontar_2023, chen_2023}, which could lead to fitting an overly dense Newkirk model. Another potential explanation is that there could be a stealth CME that pushed the coronal magnetic field outward, allowing the plasma to follow it to be perceived as having a higher density than expected, and there was not enough time for the magnetic field relaxation to occur (private communication with J. Magdalenić). However, scattering alone does not seem to fully explain the large density discrepancy. While further investigation is certainly needed regarding scattering and propagation effects on the radio waves, it is interesting to report this significant discrepancy between the model and observations, as it highlights limitations in the current modeling and suggests the need for additional physics to properly characterize the density distribution. Resolving this discrepancy could lead to important insights into the true nature of the corona.

\section{Summary and Conclusions}
\label{s_conclusions}
In this work we analysed the characteristics of a series of type III bursts that occurred on April 3$^{rd}$, 2019, during the second near-Sun encounter period of PSP. The bursts were observed in dynamic spectra taken with the PSP/FIELDS (2.6 kHz – 19 MHz) instrument, as well as in interferometric imaging with the LOFAR (20 – 80 MHz) ground-based telescope, as part of a coordinated observing campaign. The series of 16 separate weak bursts were observed over the span of $\sim$20 minutes, during an otherwise relatively quiet period. The solar disk as observed from Earth was dominated by a single active region near its centre. We combined the dynamic spectra for the LOFAR frequency range and the PSP frequency range to study the solar radio emissions within the frequency range of 2.6 kHz – 80 MHz. 

For the study, we developed a semi-automated pipeline, which allowed us to obtain the exact times and frequencies of the bursts. These we used to align the PSP to the LOFAR observations, and to generate interferometric images between 20 and 80 MHz. We performed data pre-processing of the PSP and LOFAR dynamic spectra to resample and shift the data based on the relative location of the spacecraft with respect to the Sun and Earth, and found an excellent temporal match between the two sets of observations. Thus we automatically traced the type III bursts in the dynamic spectra algorithmically and estimated frequency drift and the electron beam speeds. We found that frequency drifts remained relatively uniform between the high-frequency (LOFAR) and low-frequency (PSP) observations, as well as among the bursts, suggesting that they are related.

In addition, we imaged the type III emission at multiple frequency bands using the interferometric observations from LOFAR to determine the locations of the sources in the solar corona. The type III emissions observed were all found to occur in the same general region off the southeast limb of the Sun, leading us to conclude that they shared a single source of electron beams low in the corona.
The potential origins of these emissions are varied and include:
\begin{itemize}
    \item small-scale impulsive events such as nano-flares \citep{Ishikawa_2017, Che_2018, Chhabra_2021};
    \item plasma upflows from the active region \citep{harra2021active};
    \item coronal closed-loop structures \citep{Wu_2002};
    \item electron beams accelerated from interchange reconnection \citep{gopalswamy_2022};
    \item  high-frequency Alfven waves and/or magnetic reconnection in the outer corona \citep{morton_2015, khaled_2022}.
\end{itemize}

Our magnetic extrapolation shows that there is no open potential field to either AR12737 or AR12738, which is consistent with \citet{Cattell_2021}. Our findings are in line with the conclusions of \citet{harra2021active}, who proposed that the likely origin of these type III bursts is the AR12737 region. The type III radio bursts in \citet{harra2021active} occurred between April 1$^{st}$ and 4$^{th}$, align in time with the emergence of AR12737 near the eastern limb of the solar disk.

While potential field source surface models provide valuable insight into the large-scale magnetic topology, their reliability decreases near active regions where the field can deviate significantly from a potential configuration. Therefore, the lack of open field connectivity directly to AR12737 suggested by the PFSS model should be viewed with some caution.

This work complements those results by locating precisely the burst sources in the middle corona. We used the Newkirk density model to estimate the height of the radio sources from the Sun of one of the type III bursts, as representative of all. Combining this with PFSS magnetic modeling, we found good agreement between the centroids of the radio sources and the location of the southern open field lines in the corona, which would be required to produce radio emissions at interplanetary wavelengths in general. On the other hand, this location does not seem to be well connected to the AR itself, according to the PFSS model.

We attempted to correct the radial distance of the radio sources from the Sun by replacing the Newkirk model with more realistic MHD results from the MAS model, but we found that there is a significant discrepancy between the Newkirk model profile fitted to the observations and the MAS density.
This could result from scattering lensing the apparent burst location to a higher altitude, thus, overestimating the height of radio sources in the corona.
The presence of type III radio sources at relatively high distances in the corona, with plasma density higher than expected from the MAS model, suggests that there may be missing information in the modeling. One possibility is the existence of a stealth CME that pushed the coronal magnetic field outward, causing the plasma to appear denser than expected (see \citet{dumbovic20212019}) — or other non-obvious changes in large-scale coronal magnetic topology.
These findings demonstrate that scattering and propagation effects play a significant role in determining the location and directionality of solar radio bursts \citep{kontar_2019, kontar_2023, chen_2023}. Therefore, the discrepancy between the observed and modeled density profiles could potentially be attributed to scattering and lensing effects that make the radio sources appear higher in the corona than their true location. Further investigation is required to disentangle these effects from limitations in the density models themselves. Overall, accounting for scattering and refraction will likely lead to improved modeling of the corona and solar radio bursts.
In future work, we will also employ the Time Delay of Arrival (TDoA) technique \citep{zhang2019forward} to estimate the radio burst source positions from multi-instrument observations and compare that with the current methodology in this paper. Solar Orbiter observations shall also be included.

High-fidelity interferometric radio imaging in metric-decametric wavelengths provides a powerful method to characterise solar eruptive events. It is also becoming increasingly important for studying relatively quiet periods, during which there may be elevated levels of in situ particle fluxes. The ability to observe and image faint radio bursts such as those presented in this work, which may be related to episodes of reconnection on the solar surface, and potentially to episodes of solar wind release, is a testament to LOFAR’s power as a space weather instrument. In future work, we will automate and use our method for studying hundreds of faint bursts observed with LOFAR, and will investigate their relation to small-scale activity on the solar surface.

Through a novel combination between the LOFAR imaging and MAS model results, we observed that the type III radio bursts experienced a weakening background magnetic field, decreasing solar wind dynamic pressure and Alfven speed, increasing plasma beta and coronal temperature, and plasma rarefaction.
The radio sources appeared at larger radial distances than the models predicted, which suggests scattering and density fluctuations are important to interpreting the true burst trajectory.
The discrepancies between the observed and modeled radial distances of the radio sources suggest refinements are needed in the models to fully explain the radio imaging and modeling results. Overall, comparing the LOFAR imaging and MAS modeling for these type III bursts motivates further analysis on additional radio bursts to better understand the physical conditions that influence the propagation of radio emissions in the corona.
\begin{acknowledgements}
We thank the anonymous referee for the constructive feedback. Thanks go to Bing Ma, Marc Pulupa, Dejin Wu, and Jon Vandegriff for helping with the PSP data. We thank Pete Riley for helping with the psipy tool, and Jan Gieseler for helping with the Solar-MACH tool. Many thanks to N. Gopalswamy, L. Harra, Nour E. Raouafi, Nariaki V. Nitta, J. Magdalenic, and K. Alielden for the valuable discussions during private communications. We thank the Sunpy and Helionauts communities for the technical support. This work was supported by the Bulgarian National Science Fund, VIHREN program, under contract KP-06-DV-8/18.12.2019 (MOSAIICS project). It was also supported by the STELLAR project, funded by the European Union’s Horizon 2020 research and innovation program under grant agreement No. 952439. The authors acknowledge data usage from LOFAR core and remote stations, from the PSP/FIELDS, STEREO/SWAVES, Wind/WAVES instruments, as well as from the SDO/AIA, HMI, and EVE instruments. This research used version 4.1.5 \citep{sunpy_community2020} of the SunPy open source software package.
\end{acknowledgements}
\bibliographystyle{aa}
\bibliography{refs}
\begin{appendix}
\section{Persistent Imaging Technique}
\label{append_a}
Persistent imaging is a technique used in medical imaging, particularly ultrasound imaging, to create a continuous, real-time display of the anatomy being imaged (see \citep{pysz2011assessment} and references within). The core idea of persistent imaging is to use persistence, or the ability of the human eye to retain an image for a brief moment after it has disappeared, to create a more informative and visually clear image \citep{fredkin1995persistence, thompson2016persistence}.

At every image in a time-ordered series, the technique keeps the old pixel value if it is brighter than the current pixel value, else it takes the current pixel's value. The result is saved as the current persistence image. Then, the next image in the series is evaluated by comparing it pixel by pixel with respect to the previous persistence image. The resulting image emphasizes the changes between the current image and the previous persistent image, making them more visible to the human eye.

The persistent imaging technique can be described mathematically by a set of equations. Let $I(t,x,y)$ be the intensity at time $t$ and pixel coordinates $(x,y)$, and let $P(t,x,y)$ be the persistence image at time $t$ and pixel coordinates $(x,y)$. The persistence image at time $t$ is computed as:
\begin{equation}
    P(t,x,y) = max\{I(t,x,y), P(t-1,x,y)\}
\end{equation}
where \textbf{max} represents the maximum of its two arguments. The current image at time $t$ is then evaluated with respect to the previous persistence image as follows:
\begin{equation}
    I^`(t,x,y) = max\{I(t,x,y) - P(t-1,x,y), 0\}
\end{equation}
The resulting image $I'(t,x,y)$ is a modified version of the current image that emphasizes the differences from the previous persistence image.

The persistent imaging technique has been shown to improve the visual quality of ultrasound images and other medical imaging modalities, and is commonly used in clinical practice. In this paper, we utilize the persistent imaging technique to improve the visualization of the solar radio sources of type III emissions (Fig.~\ref{fig_persistence}).

\section{Resolving the Radio Emission Location Ambiguity}
\label{append_b}
\begin{figure}[h!]
\centering
\includegraphics[width=0.8\hsize]{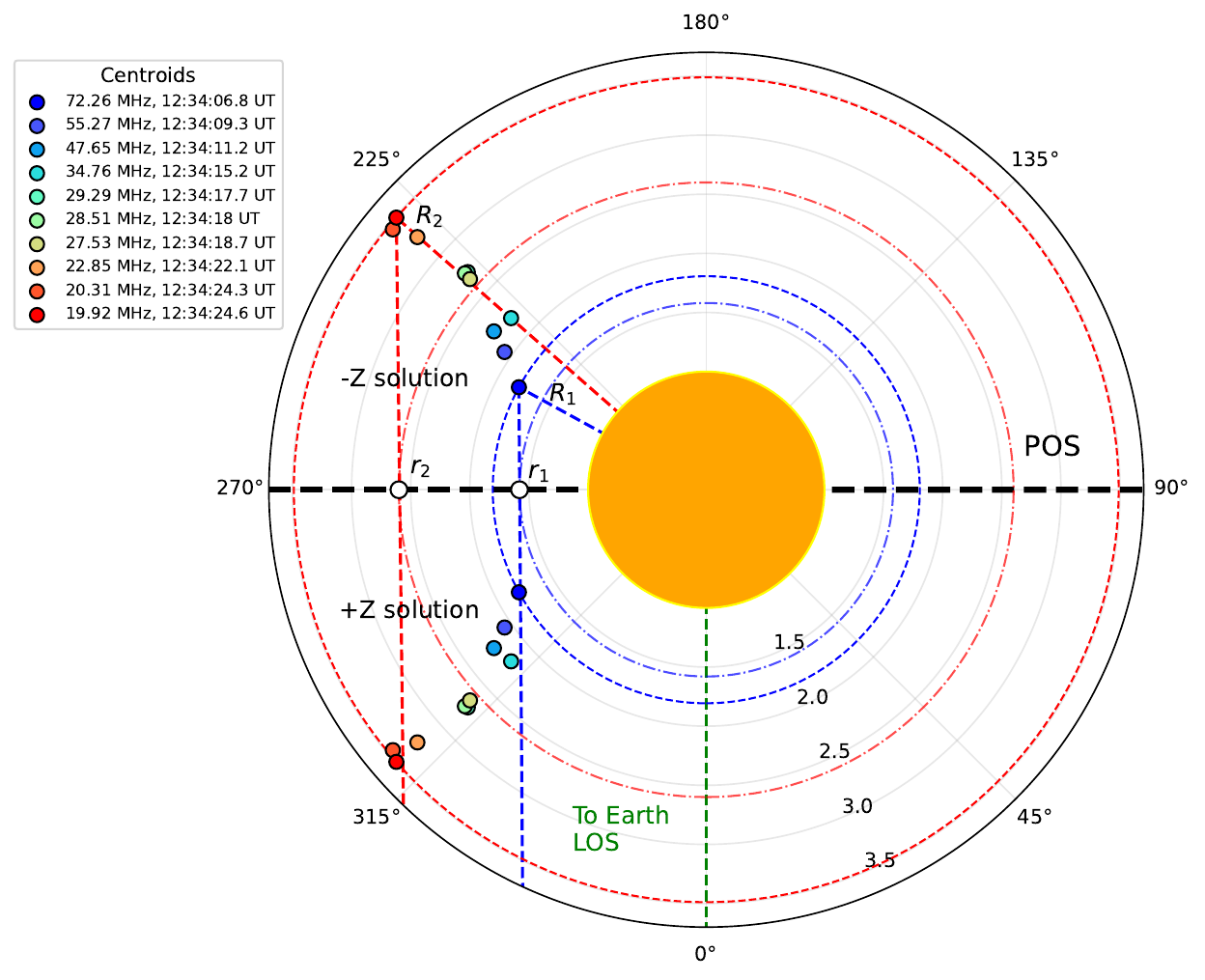}
\caption{Schematic shows the locations of the radio sources for the +Z and -Z solutions of Equation~\ref{zpos}. The Sun is located in the middle as an orange circle, with a horizontal dashed black line representing the plane of sky (POS). The vertical dashed green line represents the Sun-Earth line of sight (LOS). The dashed blue and red circles represent the plasma spheres of density equivalent to the observation frequencies of the innermost and outermost radio sources at $R_1$ and $R_2$, respectively, under the Newkirk model assumption of spherically-symmetric density distribution. The impact parameters $r_1$ and $r_2$ are the projection of $R_1$ and $R_2$ on the POS. The dot-dashed blue and red circles are the circles passing through the impact parameters $r_1$ and $r_2$, respectively.}
\label{fig_negZ_posZ}
\end{figure}

In this part, we show that the -Z solution of Equation~\ref{zpos} is highly unlikely in our case. Figure~\ref{fig_negZ_posZ} shows the positive and negative solutions of Equation~\ref{zpos}. We take the innermost and outermost coronal radio sources at $R_1$ and $R_2$, respectively, as an example. $r_1$ and $r_2$ are the projections of $R_1$ and $R_2$ on the plane of sky (POS), respectively. Harmonic radio emission from $R_1$ will theoretically be absorbed by a region along the line of sight (LOS) with plasma frequency (and corresponding density) equal to or higher than the harmonic emission frequency at $R_1$. In the case of the spherically symmetric Newkirk model, the highest density location the emission from $R_1$ could pass through is $r_1$ on the POS. Thus, for harmonic radio emission from behind the POS (-Z, where Z = 0 is defined at the center of the Sun and positive Z is towards the observer) to be observed at the Earth, it must satisfy the following condition:
\begin{equation}
    2f_{R_1} > f_{r_1}
    \label{eq_assumption}
\end{equation}
where $f_{R_1}$ is the plasma frequency of radio emission that occurred behind the POS, and $f_{r_1}$ is the plasma frequency at the projected location of $r_1$ on the POS.
The relation between the local plasma frequency and the electron density is defined by the equation
\begin{equation}
    f[MHz] = 8.93 \times 10^{-3} \sqrt{n[cm^{-3}]}
    \label{eq_plasmafreq}
\end{equation}
The Newkirk electron-density model \citep{newkirk61, newkirk67} describes the typical densities in the outer part of the corona according to the following equation
\begin{equation}
    n[cm^{-3}] = \alpha \; 4.2 \times 10^4 \; 10^{4.32 \frac{R_\odot}{r}}
    \label{eq_newkirk}
\end{equation}
where $\alpha$ is the fold number (i.e., a multiplicative factor that accounts for the density variations based on the degree of solar activity), and $r$ is the radial distance from the Sun in solar radii.
By substituting Equations~\ref{eq_plasmafreq} and~\ref{eq_newkirk} into Equation~\ref{eq_assumption}, we obtain
\begin{equation}
    \frac{n_{r_1}}{n_{R_1}} = \frac{10^{4.32 \frac{R_\odot}{r_1}}}{10^{4.32 \frac{R_\odot}{R_1}}} < 4.
\label{eq_condition}
\end{equation}
After reduction we obtain the final formula that must be satisfied under these assumptions in order for radio emission behind the POS to pass through the corona and reach the Earth
\begin{equation}
    \frac{r_1}{R_\odot} < \left( \frac{log 2}{2.16} + \frac{R_\odot}{R_1} \right)^{-1}.
\end{equation}

From Figure~\ref{fig_negZ_posZ}, $r_1$ and $r_2$ will always be smaller than $R_1$ and $R_2$, respectively. The Newkirk model requires that the density at $r_1$ and $r_2$ be significantly higher than the density at $R_1$ and $R_2$, respectively (Table~\ref{table_negZ}). Additionally, from the geometric representation in Figure~\ref{fig_negZ_posZ}, we find that the electron density at $r_1$ is higher than at ${R_1}$, hence the radio emission cannot reach the Earth from that point behind the POS \citep{Mann_2018}.

From Table~\ref{table_negZ}, the assumption of Equation~\ref{eq_condition} is not satisfied. Thus, the -Z solution is invalid in our case. This implies that the harmonic emission from behind the POS will not reach the Earth. Thus, the +Z assumption is the valid solution.
\begin{table}[h!]
\centering
\caption{Radial distances and densities at the first ($R_1$) and last ($R_2$) radio sources were obtained from the 2.5$\times$Newkirk model, as well as their impact parameters $r_1$ and $r_2$, respectively.}
\label{table_negZ}
\begin{tabular}{cccc}
\hline
Point & Radial distance ($R_\odot$) & Density (cm$^{-3}$) & Ratio ($n_r/n_R$)\\ \hline
$r_1$ & 1.58 & 5.69$\times10^7$ & \multirow{2}{*}{11.81}\\
$R_1$ & 1.81 & 4.82$\times10^6$ & \\ \hline
$r_2$ & 2.6 & 2.59$\times10^7$ & \multirow{2}{*}{14.23}\\
$R_2$ & 3.49 & 1.82$\times10^6$ & \\ \hline
\end{tabular}
\end{table}
\begin{figure}
\centering
\includegraphics[width=0.9\hsize]{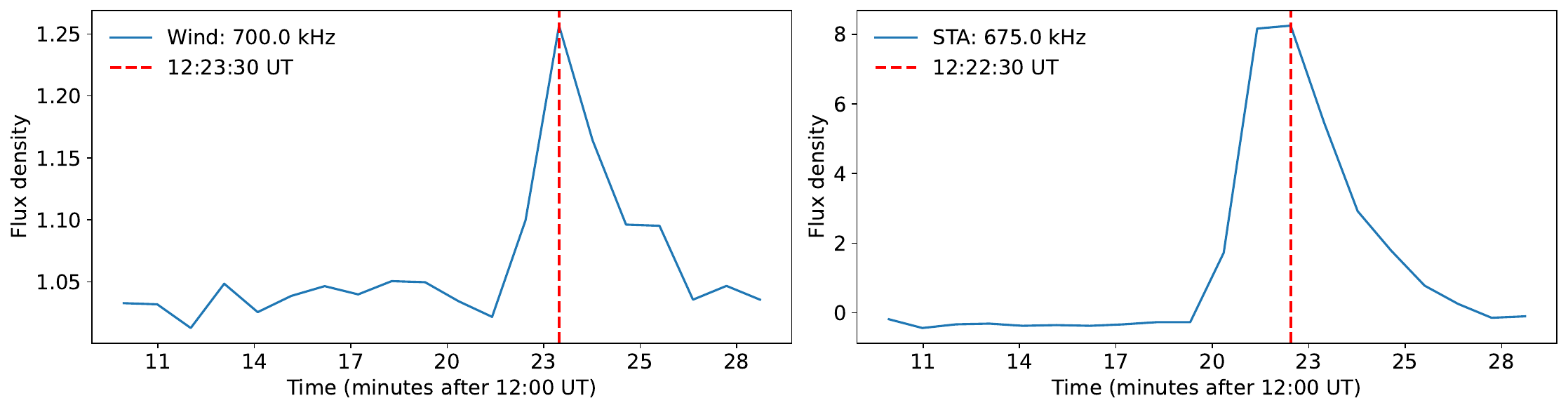}
\caption{Cut of the flux density at 700 kHz observed by Wind (left panel) and STEREO-A (right panel). For STEREO-A, there is no exact frequency channel at 700 kHz, therefore we selected the nearest one (675 kHz).}
\label{fig_cutpower}
\end{figure}

Furthermore, we analyzed the time difference of arrival of the radio emission at interplanetary wavelengths in Figure~\ref{fig_cutpower}. Specifically, we compared the timing of peak signals at a low frequency between two spacecraft, Wind and STEREO. This analysis was conducted under the assumption of two possible scenarios:
\begin{itemize}
    \item one in which the radio emission source follows a trajectory roughly equidistant between Wind and STEREO, if the $+Z$ assumption is true, or
    \item the trajectory implies significantly longer travel times from the source to Wind compared to STEREO, if the $-Z$ assumption is true.
\end{itemize}
Examining the data, we selected the frequency channel 700 kHz observed by Wind and its nearest counterpart 675 kHz for STEREO. Interestingly, the difference in the arrival times of these signals was merely one minute, within the time resolution of the instrument. This negligible difference in arrival times supports the $+Z$ assumption for the beam trajectory, meaning it travels approximately equidistant between the two spacecraft.
\end{appendix}
\end{document}